\documentclass[preprint]{aastex}

\usepackage{graphicx}

\newcommand{\ec}{$\eta$~Car}
\newcommand{\degree}{\ensuremath{^\circ}}

\begin{document}

\title{A Sea Change in Eta Carinae\altaffilmark{1,2}}

\author{Andrea Mehner\altaffilmark{3}, 
        Kris Davidson\altaffilmark{3}, 
        Roberta M.\ Humphreys\altaffilmark{3}, 
        John C.\ Martin\altaffilmark{4}, 
        Kazunori Ishibashi\altaffilmark{5}, 
        Gary J.\ Ferland\altaffilmark{6}, 
	Nolan R.\ Walborn\altaffilmark{7}   }    

  \altaffiltext{1} {Based on observations made with the NASA/ESA Hubble Space Telescope. STScI is operated by the association of Universities for Research in Astronomy, Inc. under the NASA contract  NAS 5-26555.} 
  \altaffiltext{2} {Based on observations obtained at the Gemini Observatory, which is operated by the
Association of Universities for Research in Astronomy, Inc., under a cooperative agreement
with the NSF on behalf of the Gemini partnership.}
  \altaffiltext{3} {Department of Astronomy, University of Minnesota, 
       Minneapolis, MN 55455}   
  \altaffiltext{4} {University of Illinois Springfield, Springfield, 
       IL 62703}  
  \altaffiltext{5} {Department of Physics, Nagoya University, Nagoya 464-8602}
  \altaffiltext{6} {Department of Physics \& Astronomy, University of 
       Kentucky, Lexington, KY 40506}  
  \altaffiltext{7} {Space Telescope Science Institute, Baltimore, MD 21218}   

\email{mehner@astro.umn.edu}  

\begin{abstract}
  Major stellar-wind emission features in the spectrum of $\eta$ Car have 
  recently decreased by factors of order 2 relative to the continuum.  
  This is unprecedented in the modern observational record.  
  The simplest, but unproven, explanation is a rapid decrease in the 
  wind density.  
\end{abstract}


\section{Introduction}    

Today, 150 years after the close of its Great Eruption, $\eta$ Car 
has not yet returned to thermal and rotational equilibrium
\citep{2005ASPC..332....3M,2005ASPC..332..101D}.  This fact is 
important because the ``SN impostor'' phenomenon and its aftermath   
constitute a major gap in the theory of massive stars, and $\eta$ Car 
is the only example that can be studied in detail; see reviews by   
many authors in \citet{2005ASPC..332.....H}.  Its recovery has been 
{\it unsteady\/},  with unexplained photometric and spectral changes 
in the 1890s and 1940s \citep{2008AJ....135.1249H}.

This object may have entered a phase of accelerated development 12-15 
years ago.  From 1953 to the mid-1990's, ground-based `V' photometry of 
star plus ejecta brightened at a rate of 0.024 mag/yr, with brief deviations 
smaller than $\pm$0.3 mag (Fig.\ 2 in \citealt{1999AJ....118.1777D}).  
In the past decade, however, it has risen 0.6 mag above that earlier 
trend line (Fig.\ 3 in \citealt{2009A&A...493.1093F}).   The central 
star shows a more dramatic increase,  a factor of more than 3 
in UV-to-visual {\it Hubble Space Telescope (HST))\/} data since 1998  
\citep{2004AJ....127.2352M, 2006AJ....132.2717M,
2009IAUC.9094....1D}.   A decrease in the amount of circumstellar  
dust may be responsible, but that requires some change in the wind and/or 
radiation field.  Meanwhile the periodic ``spectroscopic events'' of 1998.0, 
2003.5, and 2009.0, defined in \S2 below, differed in major respects 
\citep{2005AJ....129..900D,2010AJ....139.1534R,Corcoran:2009:Online}.
Very likely the mass-loss rate has been decreasing at an inconstant 
pace, while rotational spin-up may play a role 
\citep{2008AJ....135.1249H, 2006AJ....132.2717M, 2005AJ....129..900D, 
2003ApJ...586..432S}.  

All those discussions, however, seemed to face an embarrassing 
observational contradiction.  From the first {\it HST\/} spectroscopy in 
1991 until the Space Telescope Imaging Spectrograph (STIS) failed in 2004,   
{\it $\eta$ Car's spectrum showed no major change\/} except during the 
temporary spectroscopic events.  One might have expected some sort of 
spectral evolution to accompany the rapid brightening after 1998.   

In this letter we report a novel development:  Observations in 2007--2010 
with Gemini/GMOS and {\it HST\/}/STIS reveal major spectral changes.   
They are not subtle;  evidently the wind has been altered, at least 
temporarily and perhaps for the indefinite future.

\section{Data and Analysis}    

For long-term trends we need quantitative spectra of $\eta$ Car with 
consistent instrument characteristics, sampled over at least several 
years.   Unfortunately no suitable data set exists prior to the 
{\it HST\/} observations, which began with the Faint Object Spectrograph  
(FOS) in the 1990's \citep{1995AJ....109.1784D,1999ASPC..179..107H}, 
and continued with STIS after 1997.   Here we use {\it HST\/} 
spectroscopy of the central 
star with spatial resolution better than 0.3{\arcsec}, almost free of 
contamination by nearby ejecta.  Gemini/GMOS spectra of the central 
1{\arcsec} in 2007--2010 provide valuable independent information.

Eta Car has a complex 5.54-year spectroscopic cycle, most likely 
regulated by a companion star in an eccentric orbit, as discussed by 
many authors in \citet{2005ASPC..332.....H}.  (The periodicity was 
discovered in stages by \citealt{1984A&A...137...79Z}, 
\citealt{whitelock94}, and  \citealt{1996ApJ...460L..49D}.)   
High-excitation emission 
lines temporarily vanish during periodic ``spectroscopic events,'' 
e.g., around 1998.0, 2003.5, and 2009.0, perhaps near periastron.  
The spectrum change described in this paper is more conspicuous than 
any of those events, 
and there is no strong reason to assume that it is related to the 
5.54-year cycle.  But such a linkage might exist, and in any case the 
cycle may influence any data comparison.   Therefore we compare spectra 
at corresponding phases of successive 
cycles.  Here ``phase'' is defined by $P = 2023.0$ 
days and $t_0 =$ MJD 50814.0 = J1998.00, consistent with the 
\ec\ HST Treasury Program Archive.\footnote{    
    http://etacar.umn.edu/; see comments at the end of \S2 in   
    \citet{2010ApJ...710..729M}.  }     
Phases 0.00, 1.00, and 2.00 mark the 1998.0, 2003.5, and 2009.0 
spectroscopic events.

After a five-year hiatus, STIS obtained new spectra of $\eta$ Car 
beginning in mid-2009.  Our observations in August 2009 and March 2010 
occurred at phases 2.10 and 2.20, and fortunately some STIS data had  
been obtained approximately one and two cycles earlier, at phases 1.12 
in 2004 and 0.21 in 1999.   It is also prudent to examine data sets 
taken one cycle apart during 1998--2004.  Therefore we compare spectra 
of the star at phases 0.04 vs.\ 1.03, 1.12 vs.\ 2.10, and 
0.21 vs.\ 2.20.\footnote{
    Calendar dates 1998-03-19/2003-09-22,  2004-03-07/2009-08-19, 
    and 1999-02-21/2010-03-03;  MJD 50891/52904, 53071/55062, and 
    51230/55258. }       
The 0.04/1.03 data were close to spectroscopic events but not within
them;  in most proposed orbit models they represent longitudes 
100{\degree}--140{\degree} past periastron, with star-star separations 
2--5 times larger than at  periastron.   The 0.21/2.20 
phases were well outside the events 
\citep{2010ApJ...710..729M,1538-3881-139-5-2056}.

Improved STIS data reduction techniques were developed for the 
$\eta$ Car HST Treasury Program \citep{2006hstc.conf..247D}.  
However, in early 2010 the software has not yet been adapted to 
some format changes necessary for the new data.  On the other hand 
the current STScI data pipeline could not easily be applied to some 
of the 1998--2004 data.\footnote{
  The reason was merely a lack of certain wavelength calibration 
  files which are otherwise irrelevant here. }   
Therefore we used Treasury Program methods for the 1998--2004 STIS  
data and the STScI pipeline for the 2009--2010 data.   We extracted 
one-dimensional spectra of the star with a sampling width of 
0.25{\arcsec}.  This was broader than we would have chosen if the 
Treasury Program techniques had been employed throughout, but it is 
narrow enough to exclude most of the ejecta.

In principle the use of two reduction procedures might cause illusory 
spectrum differences,  but they would be no worse than a few percent.      
The 0.25{\arcsec} extraction width amounts to 5 CCD rows, broad enough
for good agreement in the interpolation and integration steps.  
These statements are confirmed by random checks of a few 1998--2004 
spectra reduced by both methods.  We also examined semi-raw data files -- 
flat-fielded and with cosmic ray hits removed, but otherwise unprocessed 
-- and they show the same large effects as the reduced spectra 
({\S}3 below).  Our results do not depend on absolute flux calibrations 
or precise spatial sampling.

We verified and extended our findings with Gemini/GMOS observations in 
2007--2010, reduced with the Gemini IRAF package.  They sampled 
wavelengths 3600--7200 {\AA} with slit 
width 0.5{\arcsec},    see \citet{1538-3881-139-5-2056}.   


\section{Results}   

During 1991--2004, HST/FOS and HST/STIS showed no definite  secular change 
in $\eta$ Car's stellar wind spectrum.  The H$\beta$ 
equivalent width, for instance, varied only $\pm$10\% (r.m.s.) outside 
spectroscopic events \citep{2005AJ....129..900D}.    
Figure 1a illustrates the  similarity of broad wind features in two 
successive cycles before 2004.   The qualitative ground-based record from 
1900 to 1990 shows no discernible instance of a change like that 
reported below;  see many refs.\ in \citet{2008AJ....135.1249H}. 

\begin{figure}
\epsscale{0.4}  
\plotone{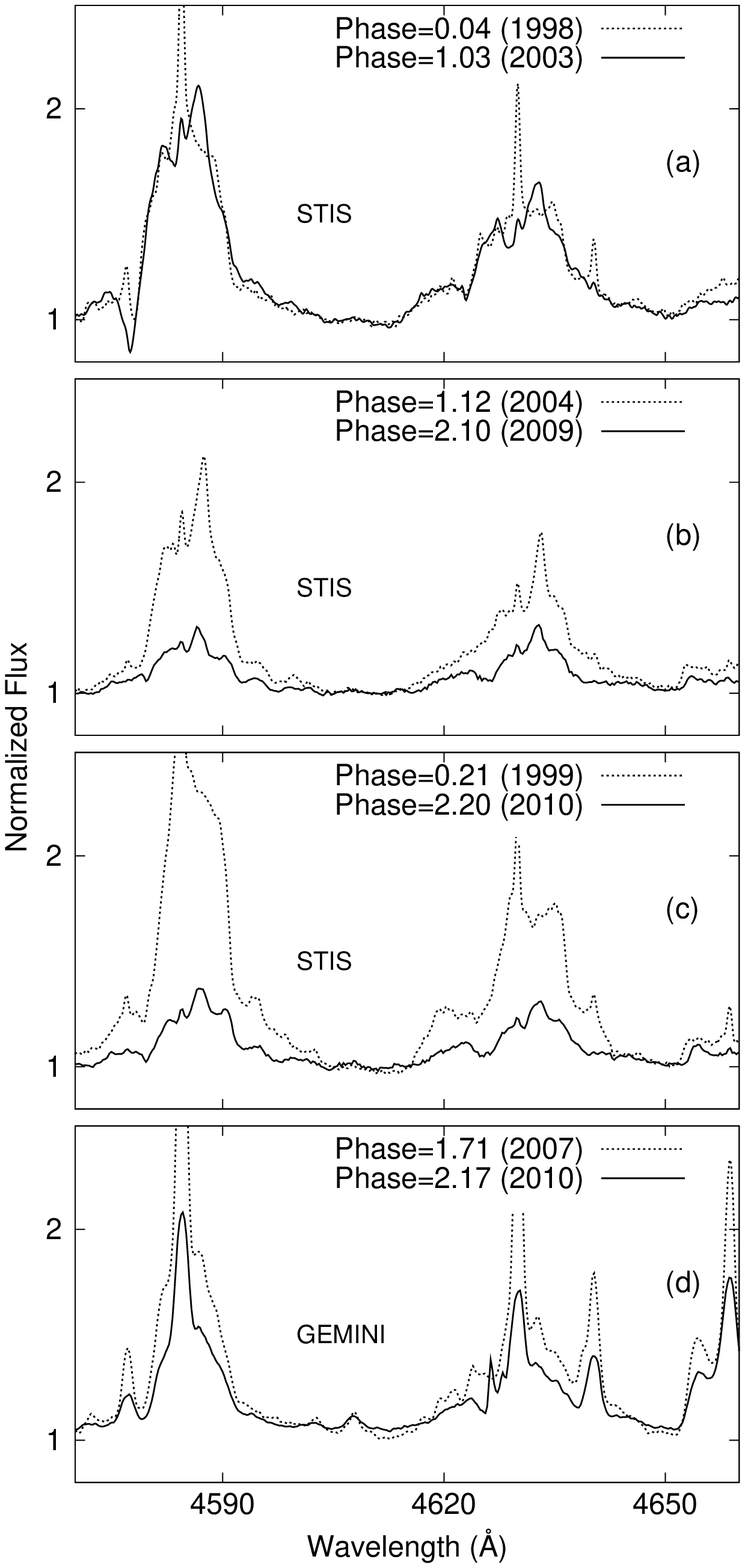}
\caption{ 
  Blends of \ion{Fe}{2}, [\ion{Fe}{2}], \ion{Cr}{2}, and [\ion{Cr}{2}] 
  near 4600 {\AA}, 1998--2010.  Flux is normalized to unity at 4605 {\AA}.    
  Panels (a,b,c) show HST/STIS data in successive 
  spectroscopic cycles (see text), while (d) shows two Gemini/GMOS spectra. 
  Spatial resolution was about 0.25{\arcsec} for STIS and 1{\arcsec} 
  for GMOS.  The {\it narrow\/} features are not  
  crucial here,  since they originate far outside the stellar wind;  their  
  decrease relative to the star may be merely an indirect consequence of 
  changes in circumstellar extinction.   The blends shown here are  
  dominated by 
  \ion{Fe}{2} $\lambda\lambda$4584.1,4585.1,4630.6, 
  [\ion{Fe}{2}] $\lambda$4641.0,  and  \ion{Cr}{2} $\lambda$4589.5. 
  \label{fig:fig1}}
\end{figure}

\begin{figure}
\epsscale{0.5}
\plotone{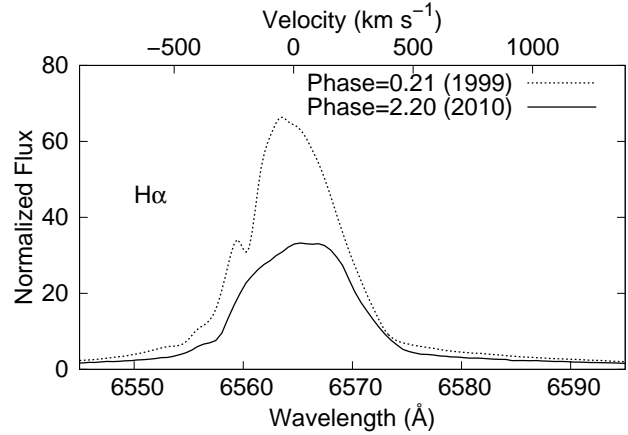}
\caption{H$\alpha$ about 400 days after the 1998 and 2009 events. Flux 
is normalized to 1.0 at 6620 \AA. Note the disappearance of external 
narrow absorption near $-145$ km s$^{-1}$.  \label{fig:fig2}}
\end{figure}

\begin{figure}
\epsscale{0.5}
\plotone{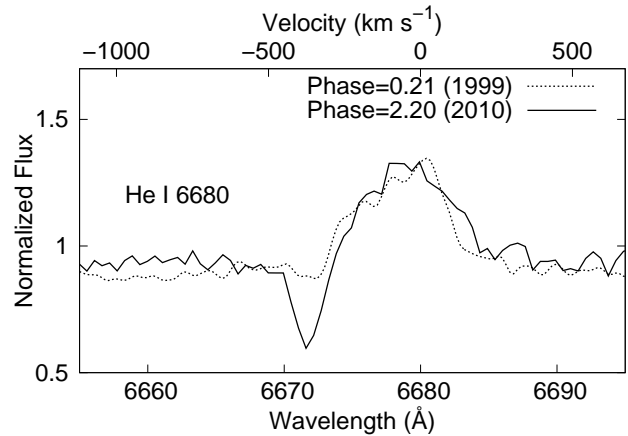}
\caption{\ion{He}{1} $\lambda$6680 about 400 days after the 1998 and 
   2009 events. Flux is normalized at 6620 \AA. The strengthened 
   absorption feature is also seen in other \ion{He}{1} lines 
   such as  $\lambda\lambda$4027,4714. \label{fig:fig3}}
\end{figure}

The 2009--2010 STIS data, however, reveal {\it the weakest broad-line 
spectrum ever seen in modern observations of this object,\/} relative 
to the underlying continuum.  We notice several effects:
\begin{enumerate}
   \item Low-excitation emission created in the stellar wind became far 
   less prominent.  For example, Fig.\ \ref{fig:fig1} shows blends of 
   \ion{Fe}{2}, [\ion{Fe}{2}], and \ion{Cr}{2} near 4600 {\AA}.   
   Phases 0.04 and 1.03 (1998 and 2003) were mutually consistent, but 
   $W_{\lambda}$ decreased by factors of 2--4 between phases 1.12 and 
   2.10 and likewise between 0.21 and 2.20 (Table 1).    
   Most of the broad lines originate in the primary star's wind, see 
   many papers and refs.\ in \citet{2005ASPC..332.....H}.
   \item The profile of H$\alpha$, the strongest emission line in the 
   violet-to-red spectrum, is altered and weakened in the recent 
   STIS data (Fig.\ \ref{fig:fig2}).  H$\alpha$ had a low flat-topped 
   profile during the 2003.5 event and then partially recovered 
   \citep{2005AJ....129..900D};  but now it is even weaker (Table 1).   
   The narrow H$\alpha$ absorption near $-145$ km s$^{-1}$  indicates  
   unusual nebular physics far outside the wind \citep{2005A&A...435..183J}.  
   Always present in 1998--2004,  this feature had weakened by 
   2007  but reappeared during the 2009.0 event 
   \citep{1984ApJ...285L..19R, 1999ASPC..179..227D, 
   2005AJ....129..900D, 1538-3881-139-5-2056, 2010AJ....139.1534R}.  
   By March 2010 it had practically vanished.    
   \item  High-excitation \ion{He}{1} emission did not weaken along with 
   the features noted above, but the P Cyg {\it absorption\/} features of 
   helium greatly strengthened after the 2009 event, 
   Fig.\ \ref{fig:fig3}.   This requires caution because \ion{He}{1} 
   varies intricately  during each cycle.  Note, however, that only 
   a few occasions in 1998--2004 showed absorption as deep as that 
   seen at phase 2.20 in March 2010;  and phase 0.21 showed practically 
   none. 
\end{enumerate} 
Table 1 lists the equivalent widths of emission and 
absorption features mentioned above.  Similar changes occurred 
throughout the violet-to-red wind spectrum.  UV emission lines around 
2600 {\AA} weakened relative to the continuum, while the overall 
brightness in that wavelength region increased by 20--30\% between 
August 2009 and March 2010.  STIS observations by other researchers 
in June 2009, covering a smaller set of wavelengths, are   
consistent with our results.\footnote{
    {\it HST\/} program 11506;  K.\ S.\ Noll, B.\ E.\ Woodgate, 
    C.\ R.\ Proffitt, \& T.\ R.\ Gull.}    

Gemini/GMOS observations in 2007--2010 confirm the reality of these 
spectrum changes, Figure 1d.   
In 2010 the GMOS data show stronger emission lines than STIS does     
(Fig.\ 1d vs.\ 1c), merely because the 1{\arcsec} ground-based 
spatial resolution allows significant contributions by ejecta far outside 
the stellar wind.  Nevertheless, equivalent widths of low-excitation 
emission blends in the GMOS data decreased by factors of about 2 between 
June 2007 and March 2010.   Most of our GMOS data at intermediate times 
were of lower quality, but they strongly suggest that the spectral 
change was progressive rather than abrupt. 


\section{What has happened?}   

We emphasize that the {\it stellar wind\/} emission lines have weakened  
relative to the continuum;  outlying ejecta will require a separate 
investigation.  The simplest explanation is a decrease in $\eta$ Car's 
primary wind density, which seems  natural for the long-term recovery 
as well as other recent data \citep{2005AJ....129..900D, 2006AJ....132.2717M, 
2008AJ....135.1249H, 2009ApJ...701L..59K, 1538-3881-139-5-2056}.  The 
surprise is in the rapidity of this development.  Long ago it was 
expected that after the year 2050 this object will appear much as it did
to Halley and Lacaille three centuries ago,  a hot fourth-magnitude star 
with a transparent rather than opaque wind \citep{1987ASSL..136..127D}. 
But now the schedule appears to be accelerated;  if the recent trend  
continues (which we cannot predict), the star will approach that goal 
in only a decade.  Even if the spectrum regresses to its earlier state,  
these developments are crucial because the 
observational record shows no precedent for them.

The effects reported in \S3 do not match the standard traits of Luminous 
Blue Variables \citep{1994PASP..106.1025H}.  The energy carried by 
$\eta$ Car's wind surpasses a bright classical LBV by a factor of 100 
or more, and its emission lines are far stronger.  When an LBV 
experiences a major eruption its  wind becomes opaque, cools below 9000 K, 
and develops a rich absorption-line spectrum within a few months.  Visual 
wavelengths brighten but the UV correspondingly fades.  Something like 
that did happen to $\eta$ Car around 1890;  but its recent record, by 
contrast, shows no perceptible  decrease in the UV/visual flux ratio.   
``LBV'' is not a very satisfactory label for this object, since 
much of its 1830--2010 behavior does not fit that category well.
If the recent change proves to be an increase rather than a decrease of 
the wind -- i.e., contrary to the hypothesis that we favor -- then 
$\eta$ Car may soon mimic a third-magnitude F-type supergiant!

Other alternatives to the decreasing-wind interpretation include, e.g., 
a change in the latitude-dependence of the wind 
\citep{2003ApJ...586..432S}, or the unusual models for $\eta$ Car favored 
by \citet{2007NewA...12..590K, 2009ApJ...701L..59K, 2009NewA...14...11K}. 
Many complications exist.  For instance, a lessened wind density 
should cause the photosphere (located in the opaque wind) to shrink 
and become hotter, eventually leading to a {\it decrease\/} 
in visual-wavelength flux.  Indeed this may have occurred in 
2006 \citep{2009A&A...493.1093F, 1538-3881-139-5-2056},  
but circumstellar dust and other factors probably dominate.

Numerous observables figure in the problem.  For example, in {\S}3 we 
mentioned that \ion{He}{1} lines have behaved differently from the 
lower-excitation features.  Helium emission and absorption processes 
in $\eta$ Car's wind depend on the companion star and have 
other special characteristics, see {\S}6 of \citet{2008AJ....135.1249H}.  
Also relevant are the 2--10 keV X-rays formed in the wind-wind collision 
zone.  \citet{2009ApJ...701L..59K} have suggested that the  
earlier-than-expected recovery of X-rays after the 2009.0 spectroscopic 
event may signal a decrease in the wind density.  Independent of that 
problem, in early 2010 the 2--10 keV flux has been about 20\% below 
the level seen in two previous 5.5-yr cycles \citep{Corcoran:2009:Online}.   
This decrease is much less extreme than the spectroscopic changes 
described in {\S}3 above;  perhaps these effects depend on 
latitude differences between our direct view of the wind 
and conditions near the wind-wind shocks \citep{2003ApJ...586..432S, 
2005ASPC..332..101D, 2008AJ....135.1249H}.   
Realistic wind models will need to be non-spherical and even 
non-axisymmetric.

Eta Car's behavior may provide spectroscopic opportunities not foreseen 
until recently.   For instance, if the wind becomes semi-transparent, 
then the temperature and radius of the primary star may become observable 
for the first time.  
Moderate-sized instruments are valuable because HST and large telescopes 
will provide, at best, only sparse temporal sampling.  Fortunately, 
ground-based observations now show $\eta$ Car -- the star itself --   
more clearly than they did ten years ago, because the diffuse  
ejecta have not brightened as fast as the star.  An obvious need 
is for {\it instrumentally homogeneous\/} series 
of spectra.  Since the wind has characteristic size scales of several 
AU and velocities of several hundred km s$^{-1}$, changes may occur on 
timescales as short as a week.

\noindent {\it Acknowledgements\/} \\
This research was partially supported by grants 11291 and 11612 from the 
Space Telescope Science Institute (STScI).  We are grateful to the staff of the Gemini South
observatory in La Serena for their help and support.


\newpage  

\begin{deluxetable}{lccccc}    
\tabletypesize{\scriptsize} 
\tablecaption{Equivalent widths of some stellar-wind emission and absorption lines measured with STIS}    
\tablewidth{0pt} 
\tablehead{ 
\colhead{Date} &
\colhead{Phase} &
 \colhead{EW (Fe II, Cr II)\tablenotemark{a}} &
 \colhead{EW (H$\alpha$)\tablenotemark{b}} & 
  \colhead{EW$_{abs}$ (He I 4714)} &
  \colhead{EW$_{abs}$ (He I 6680)} 
\\
\colhead{} &
\colhead{} &
 \colhead{(\AA)} &
 \colhead{(\AA)} & 
  \colhead{(\AA)} &
  \colhead{(\AA)} 
  }
\startdata
 1998-03-19    &   0.04     &   11.47  & 830.26 &  -0.06 &  -0.20 \\
 1999-02-21 & 0.21 & 17.79  & 899.37&  -0.10 & -0.01\\
 2003-09-22  &    1.03 &   11.03 & 614.18&   -0.11 &  -0.63\\
 2004-03-07  &   1.12  & 9.69  & 822.71&    -0.18 & -0.59\\
 2009-06-30\tablenotemark{c} & 2.08 & 3.62 & --- & -0.47 & ---\\
 2009-08-19   &  2.10    & 2.90   & 483.35      &   -0.61 &    -1.10\\
 2010-03-03 & 2.20 & 3.89 &  492.73 & -0.39 &   -0.70 
\enddata 
\tablenotetext{a}{Measured between 4570 and 4600 \AA, continuum at 4605 \AA\ and 4744 \AA.}
\tablenotetext{b}{Measured between 6510 and 6620 \AA, continuum at 6500 \AA\ and 6620 \AA.}
\tablenotetext{c}{H$\alpha$ and He I $\lambda$6680 were not observed 
   on this occasion. }
\label{tab:table1}  
\end{deluxetable}

\end{document}